\documentclass{article}
\usepackage[utf8]{inputenc} 
\usepackage{ismir,amsmath,cite,url}

\usepackage{graphicx}
\usepackage{xcolor}
\usepackage{color}
\usepackage{pbox}
\usepackage{placeins}

\usepackage{graphicx}
\usepackage{caption}
\usepackage{subcaption}
\usepackage{booktabs}
\usepackage{overpic}

\title{I've heard this before: Initial Results on TikTok's impact on the re-popularization of songs.}

\multauthor
{Breno Matos$^{1,2}$ \hspace{0.5cm} Francisco Galuppo$^{1,2}$ \hspace{0.5cm} Rennan Cordeiro$^{1,2}$ \hspace{0.5cm} Flavio Figueiredo$^1$} {
 $^1$ Universidade Federal de Minas Gerais\\
$^2$ Kunumi\\
{\tt\small \{brenomatos, franciscogaluppo, rennancordeiro, flaviovdf\} @dcc.ufmg.br}}

\def\authorname{B. Matos, F. Galuppo,  R. Cordeiro and F. Figueiredo}

\begin{document}

\maketitle
\begin{abstract}

With over a billion active users,
TikTok's video-sharing service is currently one of the largest social media websites. This rise in TikTok's popularity has made the website a central platform for music discovery. In this paper, we analyze how TikTok helps to revitalize older songs. To do so, we use both the popularity of songs shared on TikTok and how the platform allows songs to propagate to other places on the Web. We analyze data from TokBoard, a website measuring such popularity over time, and Google Trends, which captures songs' overall Web search interest. Our analysis initially focuses on whether TokBoard can cause (Granger Causality) popularity on Google Trends. Next, we examine whether TikTok and Google Trends share the same virality patterns (via a Bass Model). To our knowledge, we are one of the first works to study song re-popularization via TikTok.
\end{abstract}

\section{Introduction}\label{sec:introduction}

With the rise of novel social media websites, online music discovery becomes an ever changing ecosystem\cite{socialmediamusic}. While major players such as Spotify and YouTube Music dominate the market, short video-sharing websites like TikTok, the focus of this work, also promote music to end users. 

As stated, TikTok is mostly a video-sharing service. After a user uploads a short video, other users can use these snippets to create their videos. In TikTok, songs are present in accompanying video memes or music videos. This phenomenon is a breeding ground for audio to become viral (and thus popular). Nevertheless, many audio snippets on TikTok are from major artists and record labels. With the website becoming a platform for discovering songs from such artists, a question arises: {\em How do short video-sharing websites impact the revival of older songs?} 

\begin{figure}
    \centering
    \includegraphics[width=\linewidth]{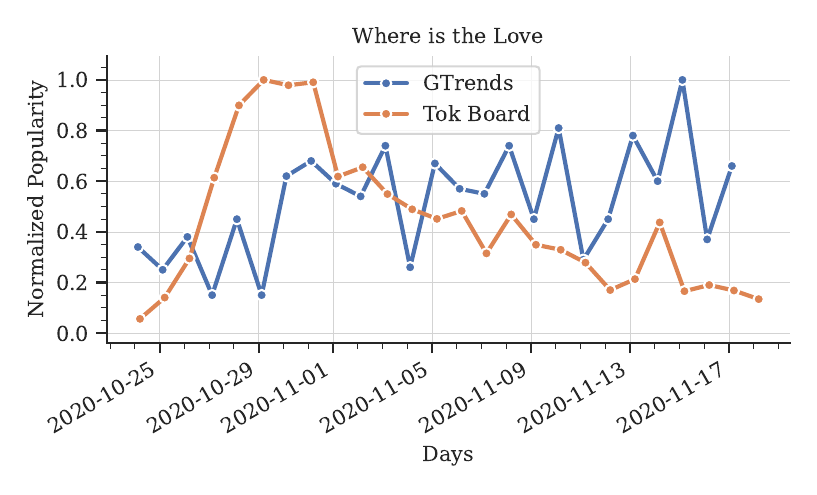}
    \caption{Popularity trend for the song "Where is the Love?" from group "Black Eyed Peas" on both Tok Board and Google Trends. The dates include the peak popularity for both platofms}
    \label{fig:bep}
\end{figure}


To tackle our research goal of understanding TikTok's impact on the re-popularization of songs, we perform an analysis of songs that were (1) released on or before September 2016 (TikTok's release date), and (2) trending on TikTok at some point in time. To define this second factor, we use data from TokBoard\footnote{tokboard.com -- A website aggregating popularity data from TikTok}. 

To motivate our work, we cite examples such as ``Dreams''\footnote{https://pitchfork.com/news/watch-mick-fleetwood-recreate-viral-fleetwood-mac-dreams-tiktok/} by {\em Fleetwood Mac} which was released in 1977, and ``Where is the Love?'' by {\em The Black Eyed Peas}\footnote{https://newsroom.tiktok.com/en-us/year-on-tiktok-music-2020}. This second example is shown in Figure~\ref{fig:bep}. As reported by news outlets, the viral trend of both songs on TikTok affected the song's overall popularity online and offline. 
This is exemplified in the figure when we see the Tok Board curve in a rise-and-fall viral trend, whereas the GTrends curve increased after the song went viral. 
Motivated by such examples, our study will mainly focus on answering the two research questions described below:

\noindent {\bf RQ1:} {\em Is it possible to predict Web search popularity based on TikTok popularity?} Here, we shall employ the
Granger Causality Test~\cite{grangercausality}, in which we evaluate if the TokBoard popularity curve can predict the Google Trends curve. If this is so, we have evidence that TikTok is causing Web search popularity. While this question sheds light on causality and prediction, we still need to understand the viral patterns of both websites. This is why we complement this question with the next one.


\noindent {\bf RQ2:} {\em Do the viral trends of TikTok transfer to Google Trends?} In this question, we shall fit TokBoard and Google Trend curves with the Bass Model~\cite{bassmodel}. The Bass Model is a simple differential equation that captures how products get adopted by a population. It has two interpretable parameters to understand virality: (1) Innovation, or who are the adopters that consume a song without influence, and (2) Immitation, who are the adopters influenced by others. After our causal analysis, we aim to understand whether TikTok trends are reflected in Web searches.

Overall, and to the best of our knowledge, ours is the first work to look into the popularity of both TikTok and Web Search based on these two aspects. We also release all code and data for this paper to foster reproducibility.\footnote{https://github.com/brenomatos/tiktok-lamir} Before discussing our dataset and results, we take the time to summarize related work in the next section. 

\section{Related Work}

Several authors have looked into popularity patterns for online music streaming services. As argued by Greenberg et al. \cite{greenberg2017music}, in current times, there is an unprecedented opportunity to investigate the musical listening habits of millions to increase our knowledge of how we consume music. 

Narrowing our focus on the impact of social media on music listening, datasets such as the million music tweets and \#nowplaying attempted to bridge Twitter and music listening habits~\cite{zangerle2014nowplaying,hauger2013million}. As with other cultural products, music is homophilic \cite{zhou2018homophily}. The authors of \cite{zhou2018homophily} argue that the usage of social media platforms can exacerbate the spread of musical content and reduce its impact. Another interesting case study was done by \cite{yeung2020did}. The authors show evidence of a growth in nostalgic music consumption during the COVID-19 pandemic \cite{yeung2020did}. Again, this shows how social media data is a powerful tool for understanding the re-popularization of songs. Complementary to the above efforts, several authors have employed social network analysis in music. Efforts ranged from looking at Classical music~\cite{bae2016scale}, to Jazz~\cite{andrade2016exploring,gleizer2003jazz}, Broadway Songs~\cite{uzzi2005collab}, as well as Popular Brazilian Music~\cite{silva2004complex,gunaratna2011mpb,andrade2020measuring}. 



With TikTok being a rather new social media platform, relatively few papers have explored its data and associated phenomena. We now summarize some of these. In \cite{shutsko2020user}, the authors perform a characterization of a thousand videos from TikTok. Overall, the authors focus on understanding the types of content shared. Complementary, \cite{bandy2020tulsaflop} analyzes TikTok's impact on collective action. Here, the authors present a case study of the \#TulsaFlop phenomenon. The hashtag was associated with a collective action focused on reducing support for a presidential candidate. Other authors propose qualitative \cite{klug2021trick,barta2021authenticity,study_behaviours,MUSICATIKTOK1,MUSICATIKTOK3} and quantitative \cite{MUSICATIKTOK2} analyses of TikTok.

\section{The TokBoard Dataset}



Since we aim to understand re-popularization, we focused on a smaller subset of the dataset, selected to reduce confounding influences present in the full dataset. Thus, we aimed to remove spurious patterns such as (1) several peaks of popularity, (2) songs not released before TikTok was created, and (3) songs where we have evidence that TikTok was the driving aspect of popularity.

\begin{figure}[t!]
    \centering
    \begin{overpic}[width=0.95\linewidth]{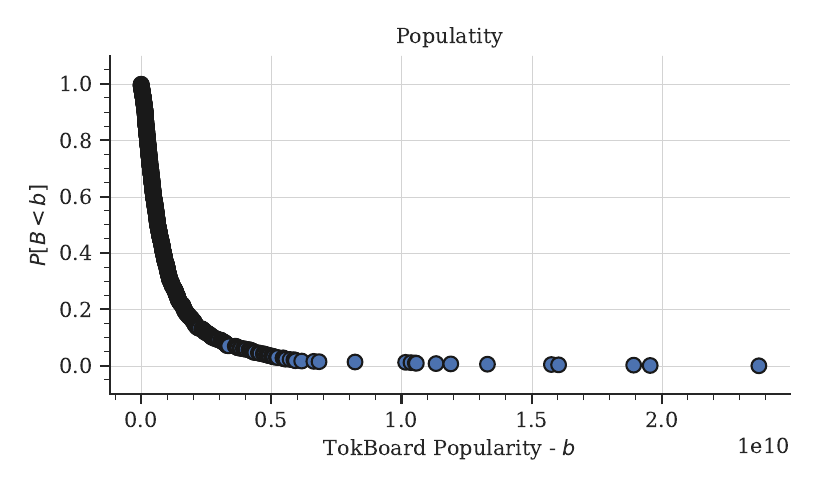}
        \put(2,23){\color{white}\rule{0.4cm}{3cm}}
        \put(3,25){\rotatebox{90}{\tiny $P(B > b)$}} 
    \end{overpic}
    \caption{Popularity of TokBoard Songs}
    \label{fig:pop}\vspace{-1em}
\end{figure}

Our dataset comes from data collected from TokBoard via a web scraper. Our research group gathered information from all 1,989 songs available in the dataset until May 19, 2022. We then queried each song’s name for Google Trends data displayed on the TokBoard website. This step captures the song’s popularity online and returns results for 1,167 songs we further filtered. To ensure comparable time-series data, we use linear interpolation on TokBoard curves to filter out any missing data. While interpolation may add minor artifacts, it maintains continuity without significant distortion.

In Figure~\ref{fig:pop}, we show the Complementary Cumulative Distribution Function (CCDF) of the popularity of TokBoard songs so far. The $y$-axis on this plot captures the fraction of songs with popularity higher than the $x$-axis. From the figure, we can see that TokBoard only monitors highly popular content (the scale of the plot is tens of billions). Indeed, the least popular song has a popularity of 11,705 streams. The most popular reaches 23,728,600,741 (over twenty three billion streams). The first, second, and third quantiles were 298,574,069, 637,920,946, and 1,380,260,724, showing that we expect popularity in the hundreds of millions or even billions of streams.

It is essential to point out that even though we collect hundreds of highly popular songs from TokBoard, not all of them will be related to our goals. For instance, not all TikTok songs will be previous hits that went viral in recent years. Moreover, songs may become popular repeatedly, exhibiting several viral-like patterns over time~\cite{figueiredo2014revisit,cheng2016cascades}.

\begin{figure*}[t!]
    \centering
    \includegraphics[width=.45\linewidth]{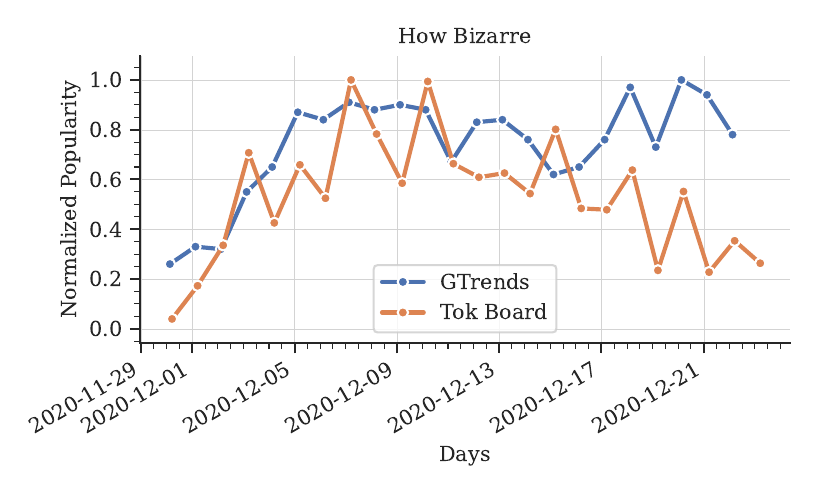}\hfill\includegraphics[width=.45\linewidth]{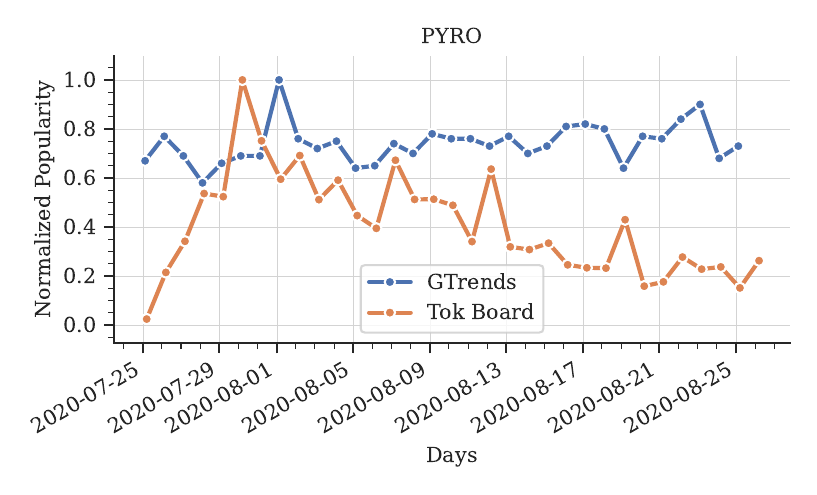} \vspace{-1em}
    \caption{Examples of Peak Focused Time Series on TokBoard} \vspace{-1em}
    \label{fig:cent}
\end{figure*}

To focus on the most prominent period of a song,  we initially filtered TokBoard popularity time series based on the peak (most popular) day. Thus, we considered points before and after the peak until the popularity of a given date was below 5\% of the total popularity. This leads to curves like the ones shown in the Introduction (see Figure~\ref{fig:bep}), and also exemplified in Figure~\ref{fig:cent}. In this second figure, we show the time series for ``How Bizarre'' by OMC (released in 1995), which was also re-popularized, and ``PYRO'' by {\em Chester Young \& Castion} (released in 2019). Although "PYRO," is popular around July 29th, 2022 (it peaks on TikTok before Google Trends), we must remove it because it was released after TikTok. Thus, it is not a proper example of re-popularization.


\begin{figure*}[t!]
\centering
\includegraphics[width=0.32\linewidth]{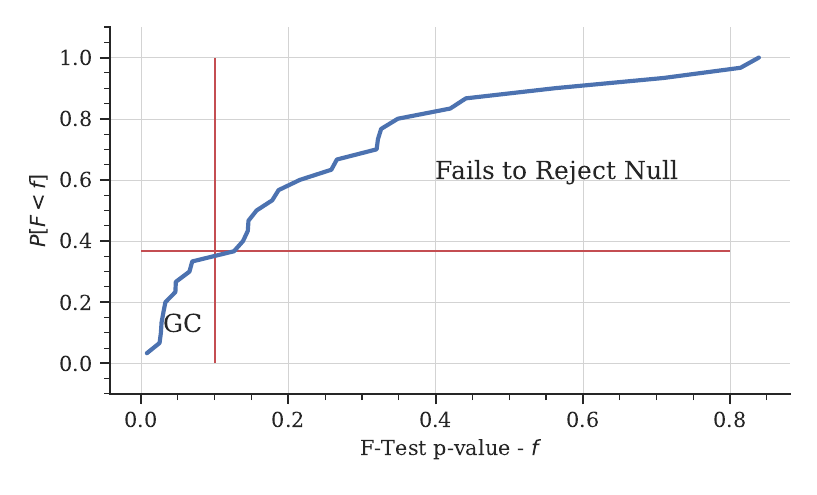}\hfill
 \includegraphics[width=0.32\linewidth]{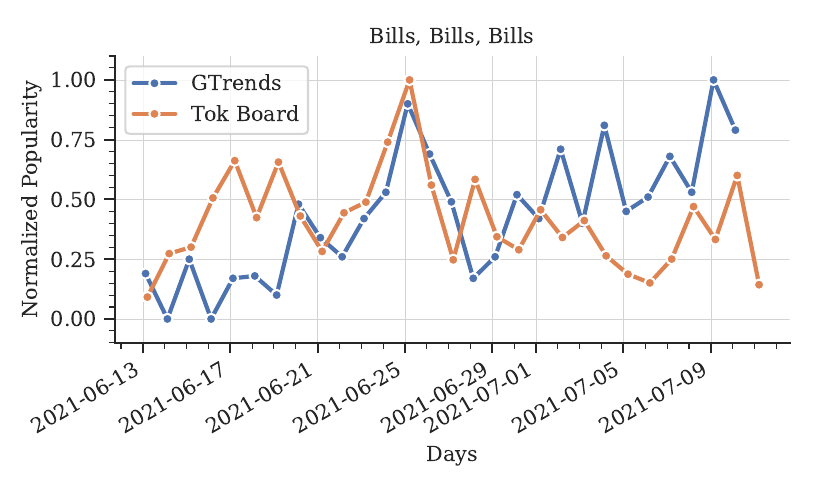}\hfill
 \includegraphics[width=0.32\linewidth]{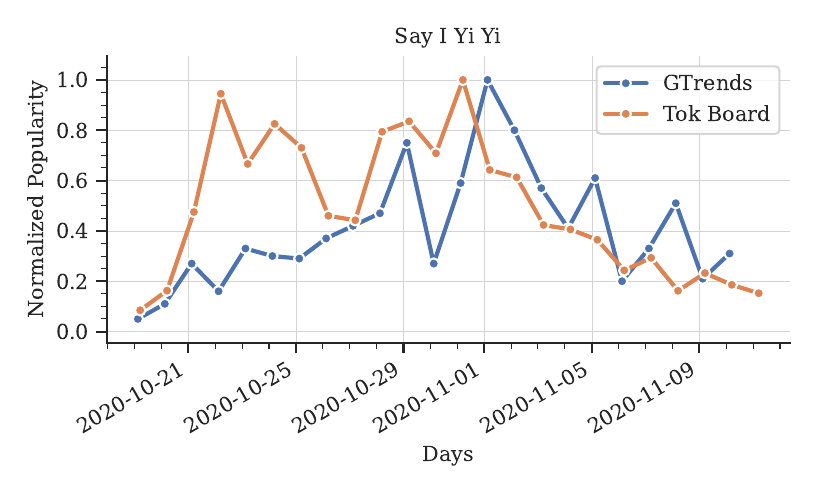}\\
 \includegraphics[width=0.32\linewidth]{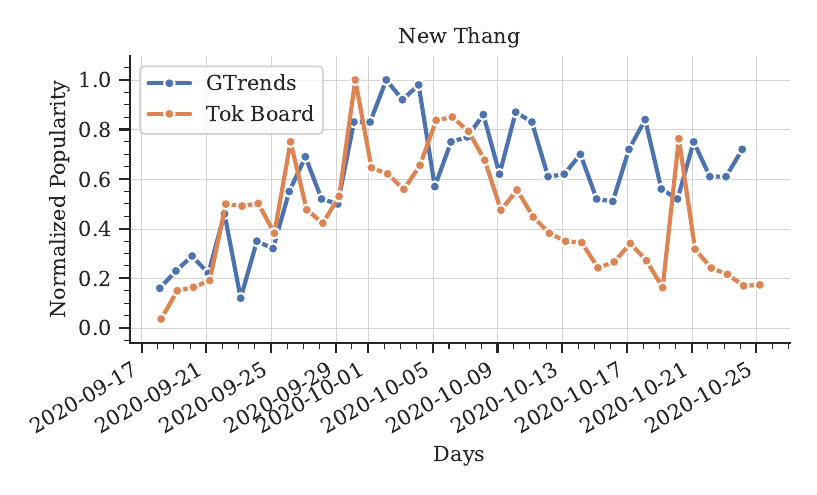}\hfill
 \includegraphics[width=0.32\linewidth]{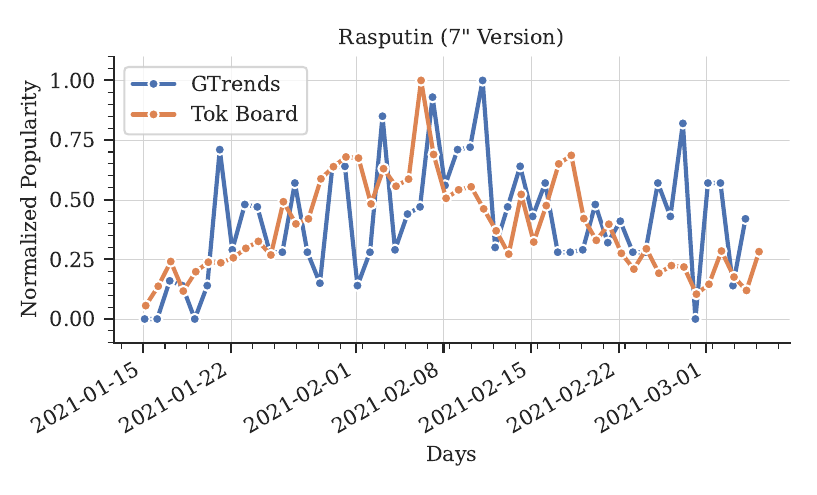}\hfill
 \includegraphics[width=0.32\linewidth]{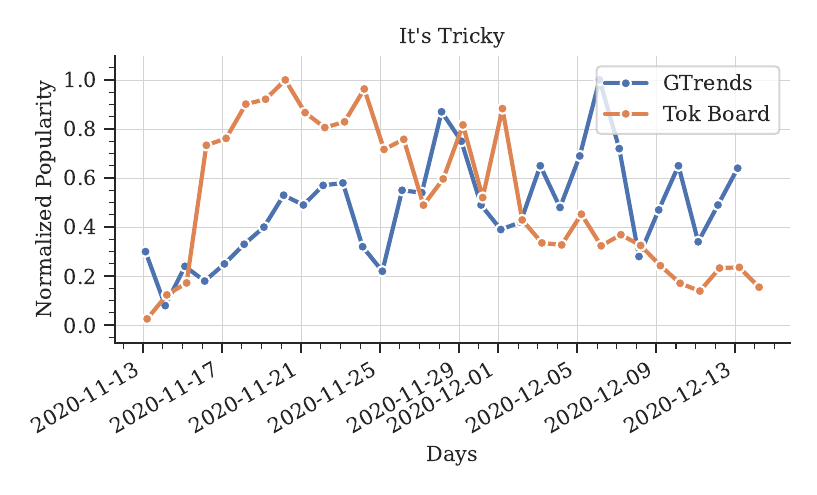} \vspace{-1em}
 \caption{P-values for Granger Causality and Some Examples of Granger Causality} \vspace{-1em}
 \label{fig:granger}
\end{figure*}


For the above reason, selected songs were released on or before September 2016, when TikTok was released worldwide. This was done in a two-fold manner. Initially, we queried MusicBrainz\footnote{https://musicbrainz.org/} for the song's title as a {\em Release Group}. We only kept songs released as {\bf Singles}; this is evidence that the artists deem it good enough to publicize. Then, we removed singles released after TikTok's birth. This left us with 85 songs. Given that MusicBrainz results may return false positives, using a Fuzzy String Matching approach\footnote{\url{https://github.com/rapidfuzz/RapidFuzz}}, we matched the MusicBrainz title and artist to the one title shown TokBoard (that combines both strings in one, e.g. "PYRO by Chester Young \& Castion"). If {\bf both} matches were above 50\% (in a fuzzy match, 50\% of the smaller string is on the largest), we kept the song. In the end, this left us with 38 songs. We manually added songs to this list subsequently, as we now detail.

\begin{figure*}[t!]
\centering
\includegraphics[width=0.32\linewidth]{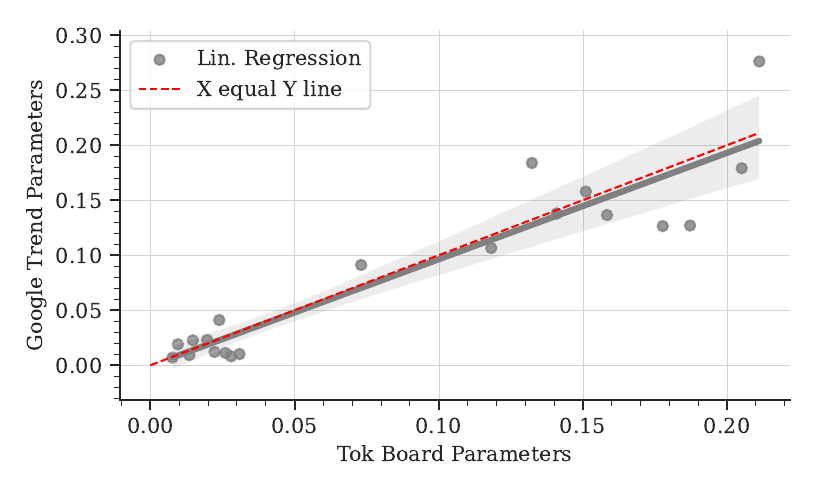}\hfill
 \includegraphics[width=0.32\linewidth]{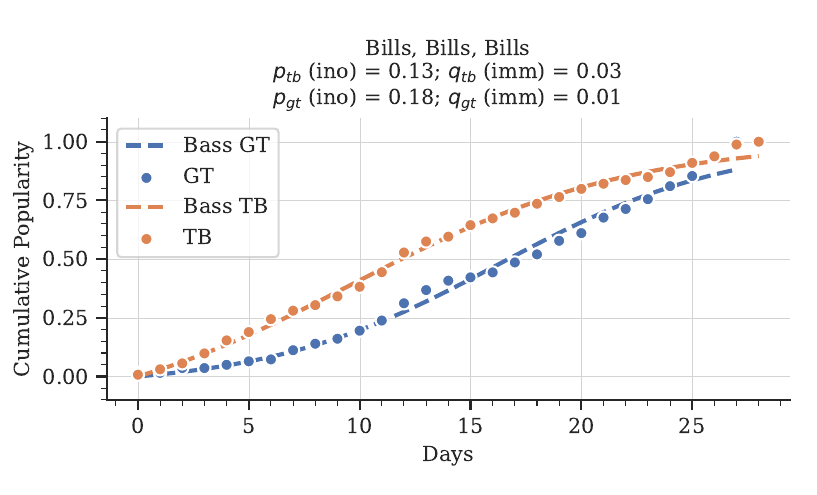}\hfill
 \includegraphics[width=0.32\linewidth]{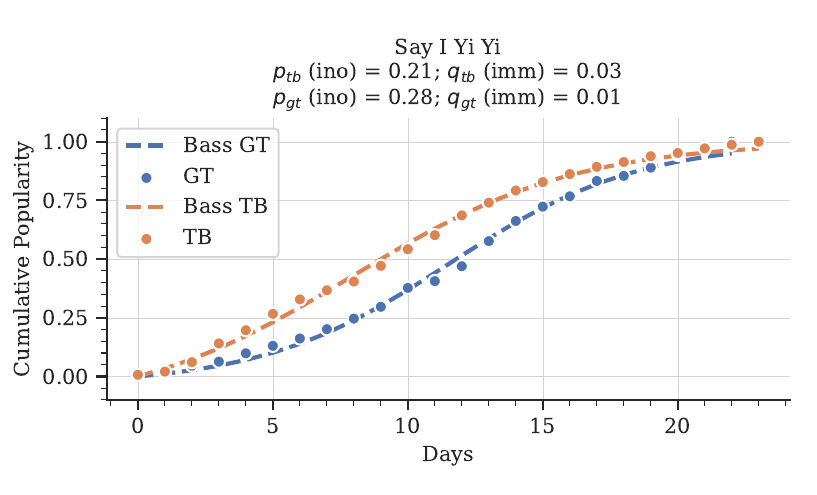}\\
 \includegraphics[width=0.32\linewidth]{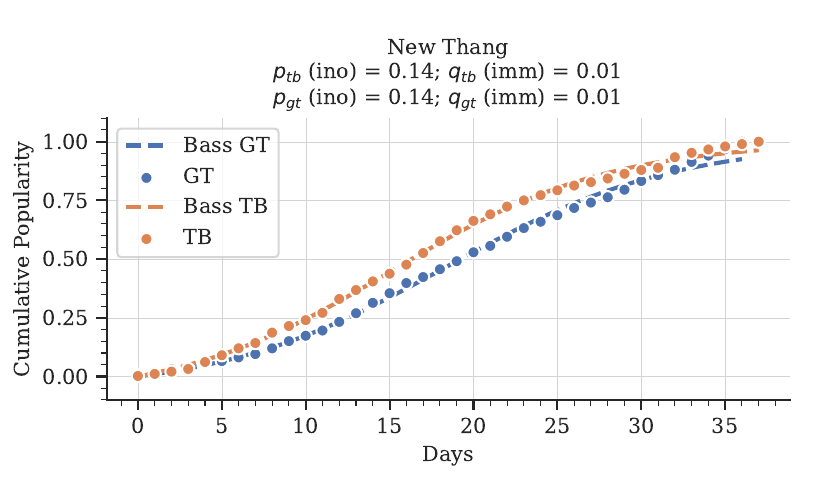}\hfill
 \includegraphics[width=0.32\linewidth]{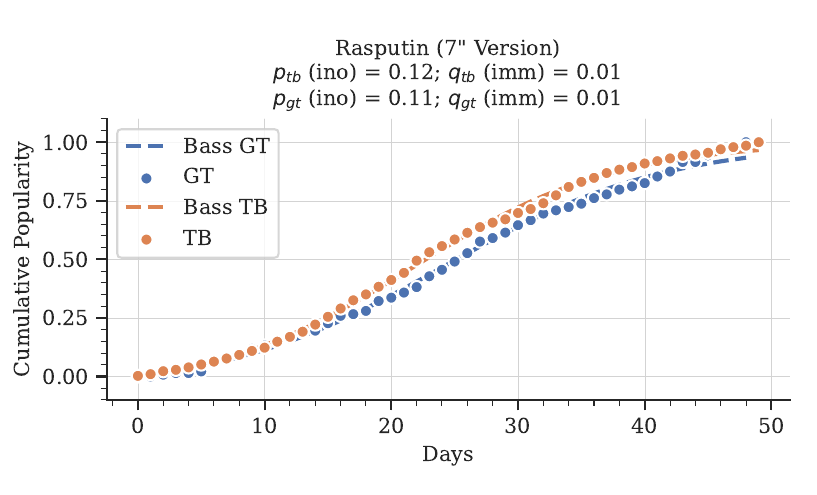}\hfill
 \includegraphics[width=0.32\linewidth]{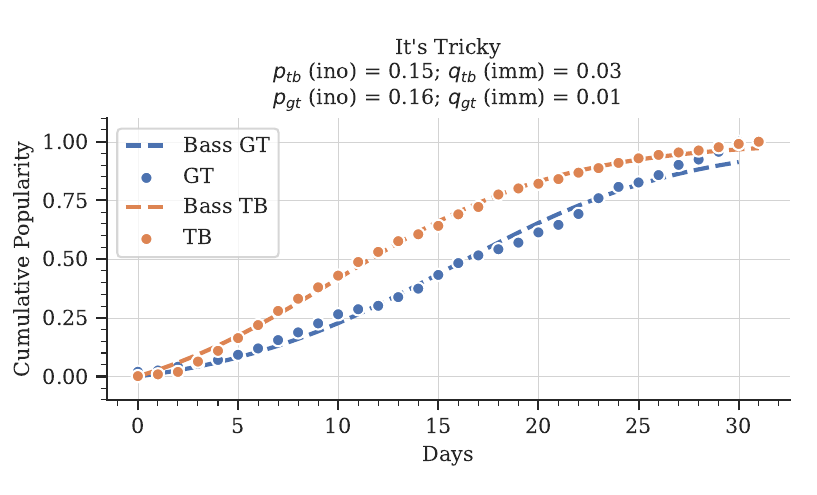}
 \caption{Scatter Plot of Bass Parameters for both websites and Some Bass Model estimation examples.}
 \label{fig:bass}
\end{figure*}
To increase the above list, we manually searched for information on each song to extract evidence that TikTok was a significant influencer on that song's popularity. To do so, we inspected results like {\em TikTok's this year on music} as well as news outlets\footnote{https://tonedeaf.thebrag.com/5-old-songs-that-tiktok-weirdly-made-famous-again/} after adding songs from this manual procedure, reaching 51 songs. Finally, we examined only songs with at least 20 data points for both time series, leaving us with 30 songs. This was done to have a minimum number of data points to execute statistical tests.

\section{Understanding the Impact of TikTok}

In this section, we showcase our results. We begin by unveiling causality using the Granger Causal test. Next, we look into whether the viral dynamics of TikTok transfer to Web searchers via a Bass model. The Granger model allows us to identify temporal correlations, while the Bass model provides insight into the diffusion and adoption patterns of songs within the digital market.



\subsection{Granger Causality}

To assess causality between time series, we employ the Granger Causality Test\cite{grangercausality}. To understand Granger causality, let $b_s(t)$ be the popularity of a song on TokBoard. Moreover, let $g_s(t)$ be the popularity of the music on Google Trends. Now, let us create two models for the popularity of Google Trends. The first will be based on data from Google Trends only, whereas the second will explore both websites. These are as follows:
\begin{align}
    g_s(t) &= \sum_{l=1}^{L} \gamma_{s,l} \, . \, g_s(t-l) + \epsilon\\
    g_s(t) &= \sum_{l=1}^{L} \gamma_{s,l} \, . \, g_s(t-l) + \sum_{l=1}^{L} \beta_{s,l} \, . \, b_s(t-l) + \epsilon
\end{align}

Here, $L$ is a fixed parameter capturing how far into the past we look. Our analysis tests several lag values for a single pair of time series. We explore $L \in [1, 5]$ in one to five days. $\gamma_{s,l}$ indicates the regression weights for the past values of Google Trends, whereas $\beta_{s,l}$ are the past values for TokBoard. Finally, $\epsilon$ is Gaussian noise, indicating that this is an ordinary linear regression model.

If, statistically speaking, model (2) is more accurate than model (1), we have evidence that the {\em past} of TikTok helps to predict the Web search popularity from Google Trends. This indicates Granger causality. We determine that this occurs based on an F-test of the sum of the squares of residuals in the models. The null hypothesis indicates a lack of evidence for causality. Thus, when rejecting it, we have evidence of causality. Figure \ref{fig:granger} displays $p$-value distribution for the 30 considered songs. Note that around 33\% of these p-values are less than $0.1$. This denotes good statistical relevance and indicates some evidence of causality for exactly ten songs. The figure also shows two examples of such songs. Overall, this result suggests that Granger Causality is less present than expected (i.e., on our dataset, TikTok is able to revitalize 33\% of songs).



\subsection{Bass Model}


Next, we employ the Bass Diffusion Model, proposed by Frank Bass \cite{doi:10.1287/mnsc.15.5.215}, to understand whether the viral aspects of songs with Granger Causality transfer across platforms. The Bass Model is defined as follows:

\begin{align}
    \frac{b_s(t)}{1 - B_s(t)} = p_s + q_s\,.\,B_s(t)
\end{align}
\noindent Here, $b_s(t)$ is the popularity of the song {\em at} $t$. Whereas, $B_s(t)$ is the popularity {\em up until}, or before, time $t$. While we use $b_s(t)$ and $B_s(t)$ for TokBoard, a similar equation is defined for Google Trends. Finally, $p_s$ and $q_s$ are estimated parameters of the model.

Parameter $p_s$ measures the innovation. Tn this case, this is the effect of new adopters. In contrast, $q_s$ measures how imitators follow the new adopters' behaviors, i.e., the influence caused by the initial adopters. Notice from the equation that $q_s$ multiplies the previous adopters; this is why it captures imitation. $p_s$ does not depend on past adopters.

We estimated the Bass Model on the ten songs with Granger Causality evidence. This was done via a Least Squares estimation of the parameters on the cumulative popularity curve. Examples are shown in Figure~\ref{fig:bass}. Note that we fit independent models for TokBoard and Google Trends. Finally, we normalized data by dividing by the sum of popularity. In this case, the cumulative curve captures the fraction of adopters or listeners. This is a common practice to fit Bass Models to data. Given that we only have 20 curve fits (two for each song), we visually inspected them. Overall, all the estimations are similar to the ones shown in Figure~\ref{fig:bass}. This result indicates that the Bass Model is a suitable approach to analyzing the adoption of TikTok songs. Overall, the results indicate that a significant portion of viral trends observed on TikTok caused an increase in web search activity, as demonstrated by the statistical findings from the Granger Causality test. In other words, the evidence suggests that TikTok plays an important role in the re-popularization of songs.

\section{Conclusions}
In this work, we present a study on TikTok's impact on popularizing songs, namely, its effect on bringing older songs back to the mainstream, employing both Granger Causality and Bass Diffusion models to examine TikTok's impact on music popularity across platforms. Our findings illustrate that TikTok significantly contributes to the re-popularization of songs by amplifying their visibility and engagement, as evidenced by the causality and virality patterns that often translate into increased web search activity. Our results showcase a robust methodology alongside a first-of-its-kind analysis, offering insights into how digital platforms can extend the lifecycle of music beyond traditional release cycles.

For future work, a potential direction is modeling the economic impact of TikTok trends on streaming and sales data. Also, we would like to explore the demographic and geographic variations in TikTok-driven music trends, which may provide deeper insights into user engagement.

\bibliography{references}

%
%
%
%
%

\end{document}